\documentclass[12pt]{iopart}

\usepackage{longtable}
\usepackage{graphicx} 
\usepackage{hyperref} \usepackage{color}

\begin{document}
\title{Valence Band Circular Dichroism in non-magnetic Ag/Ru(0001) at normal emission}
\date{\today}

\author{Arantzazu Mascaraque$^1$, T. Onur Mente\c{s}$^2$, Kevin F. McCarty$^3$, Jose F. Marco$^4$, Andreas K. Schmid$^5$, Andrea Locatelli$^2$, Juan de la Figuera$^4$}
\address{1 Universidad Complutense de Madrid, Madrid 28040, Spain}
\address{2 Sincrotrone Trieste S.C.p.A, Basovizza, Trieste 34149, Italy}
\address{3 Sandia National Laboratories, Livermore, California 94550, USA}
\address{4 Instituto de Qu\'{\i}mica-F\'{\i}sica Rocasolano, CSIC, Madrid 28006, Spain}
\address{5 Lawrence Berkeley National Laboratory, Berkeley, California 94720, USA}

\begin{abstract}

For the non-magnetic system of Ag films on Ru(0001), we have measured the circular dichroism of photoelectrons emitted along the surface normal, the geometry typically used in photoemission electron microscopy (PEEM). Photoemission spectra were acquired from micrometer-sized regions having uniformly thick Ag films on a single, atomically flat Ru terrace. For a single Ag layer, we find a circular dichroism that exceeds 6\% at the
{\em d}-derived band region around 4.5 eV binding energy. The dichroism decreases as the Ag film thickness increases to three atomic layers. We discuss the origin of the circular dichroism in terms of the symmetry lowering that can occur even in normal emission.

\end{abstract}

\maketitle

Dichroism in photoemission refers to a change in the
spectral intensity as a function of the polarization of the incident light. The technique provides insight into the detailed band structure of
surfaces and thin films \cite{KuchRPP2001,Henk1996}. In particular, circular
dichroism (CD) is often exploited to characterize the magnetic
properties of materials. Magnetic CD, especially using synchrotron light sources to excite near absorption thresholds, is a popular technique that provides element-resolved
magnetic information. X-ray magnetic dichroism
with photoemission electron microscopy, XMCD-PEEM, provides spatially resolved magnetic characterization. This approach of coupling dichroism measurements with microscopy is becoming popular  \cite{Schneider97} because of its ability to study the effects of topography, roughness, surface composition variations and magnetic domain structures. PEEM instruments usually operate by imaging photoelectrons emitted close to the surface normal. As circular dichroism is commonly attributed to either magnetism or sample chirality, understanding what conditions cause dichroism is important. Here we report substantial circular dichroism from a non-magnetic, non-chiral surface, Ag on Ru(0001), in normal emission. 
Ag on Ru(0001) has been studied by techniques including
STM \cite{HwangPRL1995,LingPRL2004}, photoemission \cite{BzowskiPRB1995}, and low-energy electron microscopy and diffraction \cite{LingSS2006}. Ag on Ru has been
used as a model system to understand how band structure evolves from a 2D system
to a 3D crystal \cite{BzowskiPRB1995}.

To understand under which circumstances circular dichroism is possible, the
symmetry of the whole experimental configuration has to be considered. In addition to the possible presence of magnetization or atomic/molecular level chirality, this includes the geometry of the experimental configuration as given by the direction of the incident light relative both to the surface normal and to the crystallographic axis of the sample, the polarization of the incident light and the direction of electron emission. Feder and Henk's rule \cite{FederHenk1996} indicates that dichroism can be observed when the two experimental configurations used for the different light polarizations are not connected by a symmetry operation of the symmetry point group of the sample.

\begin{figure}
\centerline{\includegraphics[width=0.7\textwidth]{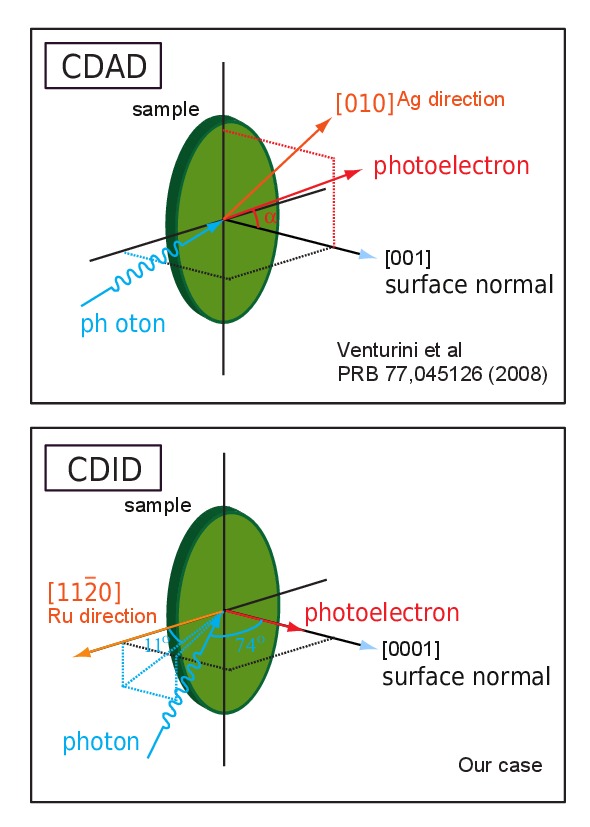}}
 \caption{Top: Circular dichroism in the angular distribution of electrons (CDAD) geometry studied in the literature where photoelectrons are detected along directions that are not normal to the surface. Bottom: Experimental configuration of our LEEM/PEEM
experiment where the electron detection direction is normal to the crystal surface. The incoming X-ray beam is oriented at a polar angle of 16$^\circ$ from the sample surface, with an azimuthal angle of 11$^\circ$ relative to a Ru $[11\overline{2}0]$ compact direction within the Ru surface.\label{experiment}}
\end{figure}

Magnetic surfaces \cite{KuchRPP2001} and surface layers of chiral molecules \cite{cherepkov_circular_1982,dubs_circular_1985} inherently generate low-symmetry conditions and circular dichroism is often observable. But circular dichroism is also possible in the absence of magnetism and molecular chirality, as observed from surfaces including Ag(001) \cite{venturini_soft_2008}, 
Pt(111) \cite{oepen_spin-dependent_1986,garbe_spin-dependent_1989,
fecher_dichroism_2002}, Pd(111) \cite{schnhense_circular_1991},
Si(001) \cite{daimon_circular_1995} and other systems \cite{fecher_gh_circular_1999}. In these studies, chiral conditions are the result of non-collinear alignments of relevant vectors including high-symmetry directions along the crystal surface and the directions of the incident light, the electron emission and the surface normal. For cases involving off-normal electron emission, this type of dichroism has been termed circular dichroism in the angular distribution of electrons (CDAD \cite{ritchie_theoretical_1975}, see top panel of figure~\ref{experiment}). Even in the higher symmetry geometry of illumination along the surface normal, CDAD can result depending on the relative alignment of the sample symmetry planes and the electron emission direction  \cite{oepen_spin-dependent_1986} and has been mapped across large ranges of off-normal
emission angles \cite{daimon_circular_1995,garbe_spin-dependent_1989}.

Chirality can also occur when collecting photoelectrons emitted along the surface normal, a geometry different from CDAD but of interest to PEEM research. In this case, low-symmetry conditions are achieved when the light is not incident along the surface normal and not aligned along a high-symmetry azimuth in the substrate plane. Compared to CDAD, non-magnetic circular dichroism involving electron emission in the surface-normal direction remains less-well explored. However, an analytical study \cite{Henk1996} indicated the possibility of observing circular dichroism due to spin-orbit coupling under non-normal light incidence and normal emission on a three-fold symmetric surface. 

In this paper we present a combined low-energy electron microscopy (LEEM) and valence-band (VB) dichroism photoemission spectromicroscopy characterization of atomically flat 1--4~ML thick Ag/Ru(0001) films. For the PEEM measurements, the incoming X-ray beam is along a direction close to in-plane but not aligned with a symmetry plane of the crystal. (The experimental geometry is described in figure~\ref{experiment}). We observe a
significant circular dichroism characterized by an asymmetry factor $A$ \cite{Note1}
that exceeds 6\% at particular electron energies. In analogy to CDAD, we term this effect circular
dichroism in the incidence distribution of photons (CDID).

The experiments were performed at the spectroscopic photoemission and low-energy electron microscope (SPELEEM) system at the Nanospectroscopy beamline of
the storage ring Elettra \cite{Schmidt1998}. The Ru(0001) single crystal preparation method has been described elsewhere \cite{ArantzazuPRB2009}. Ag was deposited from a commercial e-beam doser. The silver coverage is given in terms of the atomic layers as detected by LEEM. The valence band spectra were acquired from silver islands of a given thickness deposited on a single terrace of the substrate.
 
In PEEM mode an X-ray beam is used as the illumination source
and the photoelectrons are energy filtered by an hemispherical energy analyzer before being formed into an image. In selected-area spectroscopic
(``dispersive plane'') mode, photoelectrons from a surface region 2 $\mu$m in diameter are dispersed in energy along the exit slit of the analyzer. This line is imaged by the same two-dimensional detector used for spatial imaging with an energy resolution of 0.2\ eV. To correct for spatial variations in sensitivity, the spectra are divided by a featureless background spectrum acquired by changing the photon energy. The energy scale is measured by the position of the photoelectrons along the detector after calibrating using elastic electrons of known energy and including a non-linear correction. No
further processing of the spectra is performed.

The photoelectrons are collected in the direction normal to the film (see
figure~\ref{experiment}). The alignment of the surface normal in the instrument is better than 1$^\circ$. The angular acceptance in PEEM mode, which depends on bias applied to the sample, is 3.2$^\circ$ for the 80
eV photons used as incident beam. The sample orientation is fixed relative to the incoming X-ray beam. The angle
of incidence is 74$^\circ$ off the normal to the sample. The azimuthal
direction is 11$^\circ$ off a compact direction of the Ru(0001) surface, as
measured from the low-energy electron diffraction pattern (LEED) acquired in LEEM mode. 

No signs of Ag/Ru mixing were observed at the 290$^\circ$C growth temperature of our
experiments~\cite{LingSS2006}, in agreement with the bulk inmiscibility of Ag and Ru. In monolayer
films, the large mismatch between the Ag and Ru in-plane spacings is accommodated by forming dislocation networks located at the Ag/Ru interface \cite{HwangPRL1995,LingSS2006}. The single layer films studied here are fully dense, being in equilibrium with two-layer regions. These single layer films have a dislocation network referred to as the long-period herringbone structure  \cite{HwangPRL1995,LingSS2006}. Thicker films present a
distinctive dislocation network \cite{LingPRL2004} that efficiently relieves the
lattice mismatch with the substrate, giving rise to a silver layer of
intermediate density relative to both the Ru substrate and subsequent silver
layers: the misfit dislocations in such films are located not only at the Ag/Ru
interface but also at the 1 ML Ag/2 ML Ag interface.

\begin{figure}
\centerline{\includegraphics[width=0.7\textwidth]{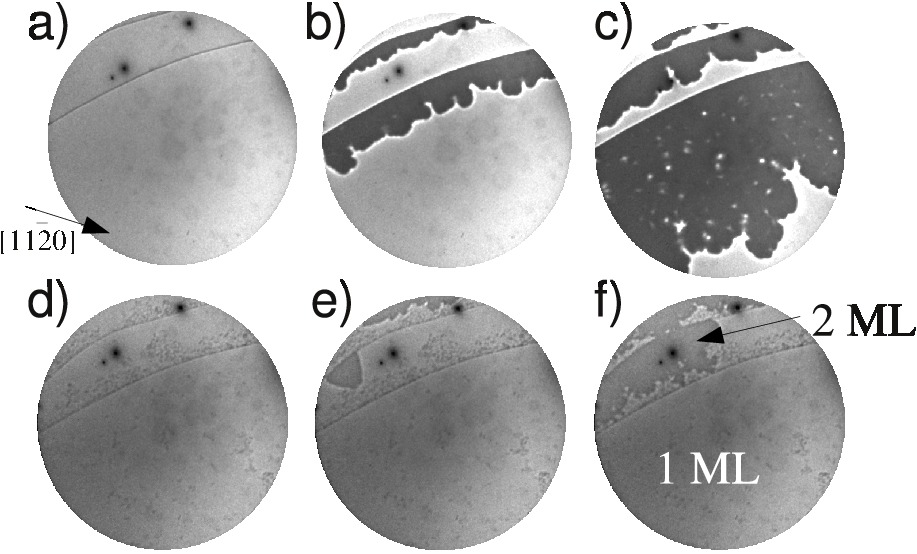}}
 \caption{Layer-by-layer growth of Ag/Ru(0001) up to the 2~ML observed in real time by
LEEM. The top row of images [a)-c)] were acquired
with an electron energy of 5.7~eV, and show the growth of a 1~ML film. The lower
row [d)-f)] were measured with an electron energy of 4.7~eV, and show the
nucleation and growth of 2~ML regions. The field of view is 10~$\mu$m. A compact direction of the Ru substrate is shown in a).\label{growth1and2ML}}
\end{figure}

In figure~\ref{growth1and2ML} LEEM images are selected from a sequence acquired during the growth of the first two Ag layers
LEEM with the temperature of the substrate kept at 290$^\circ$C. The reflectivity of each Ag layer and that
of the substrate depends on the electron energy selected. The figure shows images
acquired at energies that provide contrast in LEEM for the 1~ML vs
substrate, and 2 ML vs 1 ML thickness respectively. The growth in the large
terrace shown in figure~\ref{growth1and2ML} follows a layer-by-layer growth mode.
The growth front is more ramified for the first layer, and proceeds in step-flow
mode: there is no island nucleation on the middle of the Ru terraces. The second
layer also grows from the preexisting substrate steps but it presents a more
compact growing front that tends to be aligned along the high symmetry
directions of the substrate. 

\begin{figure}
\centerline{\includegraphics[width=0.7\textwidth]{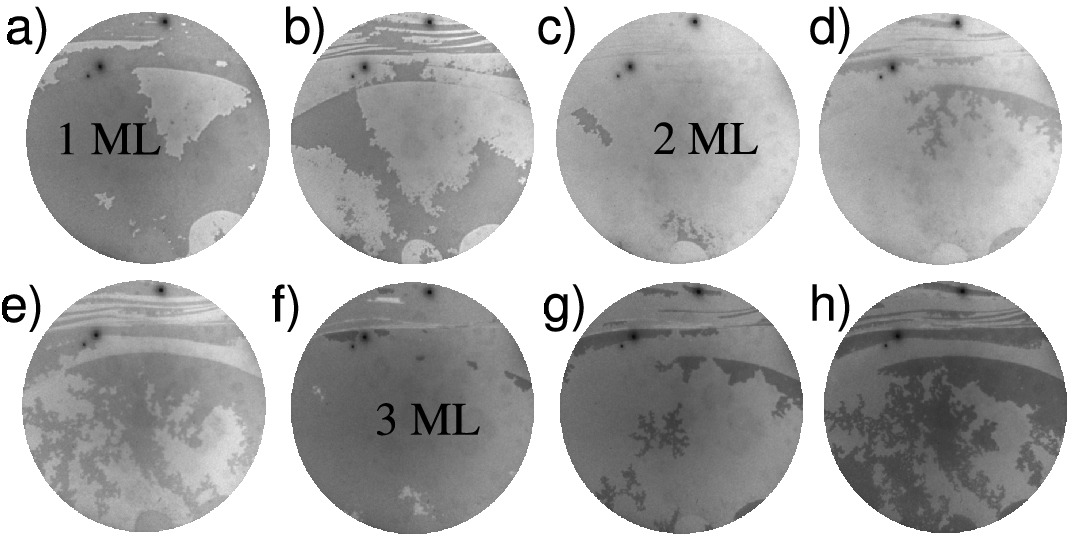}}
 \caption{a)-h) LEEM images from a sequence acquired while growing two
additional layers on top of the previous film. Regions with each film thickness are
labeled. The field of view of the images is 10~$\mu$m. \label{growth3and4}}
\end{figure}

Silver on Ru has a strong tendency to form three dimensional islands on top of
the 2~ML wetting layer,  i.e., it presents Stranski-Krastanov growth. In
particular, small substrate terraces promote 3D growth \cite{LingSS2004}.
Nevertheless, Ag/Ru(0001) can grow layer-by-layer up to at least 10 ML on large
substrate terraces taking advantage of the kinetic limitations to the nucleation
of new layers. In this way, we obtained photoemission spectra from regions having only a unique number of Ag layers. In contrast, averaging techniques
such as conventional photoemission spectroscopy cannot provide layer-by-layer information above the bilayer thickness (the wetting layer in Ag/Ru is a bilayer \cite{LingPRL2004}, so annealing a multilevel film gives rise to a bilayer film in coexistence with 3D islands). In figure~\ref{growth3and4} we present selected snapshots from a sequence of images
acquired while continuing the growth of the surface previously studied. The
growth front of the third and the fourth layers is more ramified than for the
thinner layers. Nevertheless, the growth still proceeds by step-flow, and in an
strict layer-by-layer mode.

\begin{figure}
\centerline{\includegraphics[width=0.7\textwidth]{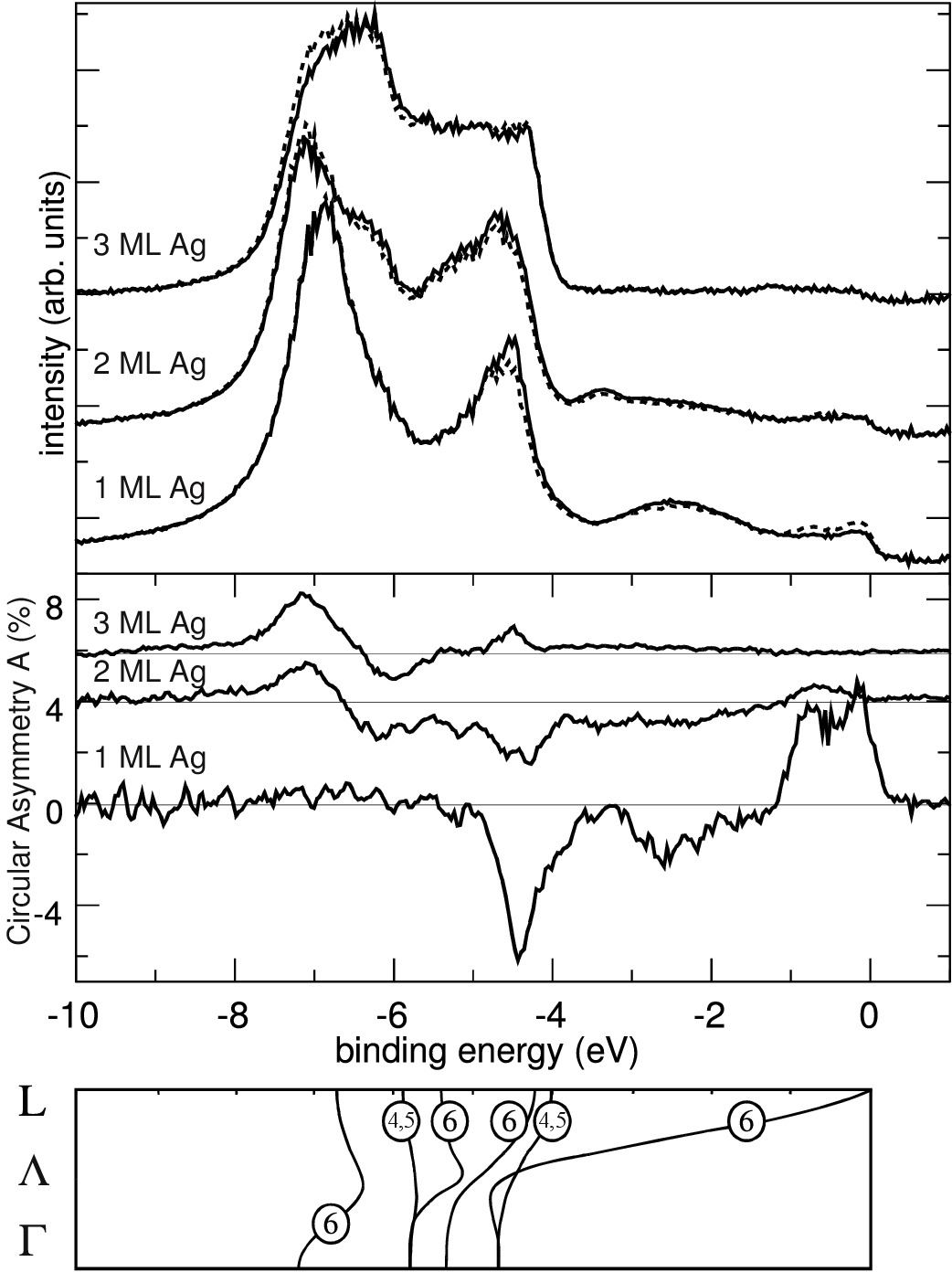}}
 \caption{Valence band dichroism observed in Ag on Ru for circular polarized
light. Top panel: photoemission spectra from 1 to 3~ML Ag regions. Positive circular polarized light spectra are shown with solid lines, while negative circular polarized light ones are shown with dots. The spectra have been offset for clarity.
Middle panel: Asymmetry \cite{Note1} factor A for the same spectra\label{circular_dichroism}.
Bottom panel: the calculated relativistic band structure of bulk Ag, from
Ref.~ \cite{EckardtJPf1984}. The labels correspond to the relativistic double group
representations.}
\end{figure}

The electronic structure of Ag/Ru(0001) has been studied for the
thickness range from submonolayer to bulk \cite{BzowskiPRB1995}. In
previous work, the growth was performed at room temperature, which is known to
produce a rough surface with several exposed thicknesses. In addition several
calculations exist for the band structure of Ag(111). In
figure~\ref{circular_dichroism} (see lower panel) we reproduce the bulk band-structure of Ag as reported in Ref.~\cite{EckardtJPf1984}. The Ag bulk
band structure has a single sp-derived band crossing from the Fermi level down
to the 4--7~eV binding energy region, where several flatter {\em d}-derived bands are
located. Ru presents a projected bandgap in the (0001)
direction for 3--6~eV binding energy. Above it, the {\em d}-derived bands are
concentrated in the 2--3~eV region.

In the upper panel of figure~\ref{circular_dichroism} we show the
photoemission spectra acquired from 2~$\mu$m-diameter regions of the Ag film. Before
acquiring the spectra, the area was explored in LEEM mode to check that each
spectra was acquired from an area with uniform Ag thickness, i.e. either one, two or three atomic layers.

There is a high intensity of photoemitted electrons with a binding energy
between 4--7~eV, assigned to the relatively flat {\em d}-derived bands of Ag. The
broad peak at 2.5~eV, clearly observed for a single monolayer of Ag, almost
disappears for the thicker films. This suggests that this spectral feature could
arise from the {\em d}-derived bands of Ru, consistent the with screening of electrons with a kinetic energy of 73~eV through the silver films. In addition, the sp-derived band of Ag is
quantized by the Ru band gap, giving rise to quantum well states in the 0--4~eV binding energy range.
The differences in the spectra of figure~\ref{circular_dichroism} with previously reported data \cite{BzowskiPRB1995} 
are attributed to the different structure and morphology of the films. In our case, the data is acquired from selected areas of a given thickness, grown at a temperature where the dislocation network present in the films is well characterized \cite{LingSS2004}.

Our experimental observation of circular dichroism is shown in
the middle panel of figure~\ref{circular_dichroism}. The asymmetry factor A is significant and clearly depends on the thickness of the Ag film. The largest asymmetry occurs in the
Ag monolayer, which has the long-period herringbone structure. For this case, the
dichroism reaches 6\% at the upper edge of the Ag d-band region at binding
energies close to 4.5~eV. There is also substantial dichroism between the Fermi level and a binding energy of 1~eV. A smaller dichroism is observed in the 2--3~eV region. In 2--3~ML
thick films, all these contributions decrease and disappear, while a small
dichroism of opposite sign appears in the 4.5~eV region. On the other hand, the dichroism increases in the 7.5~eV region.

From a simple comparison
with the relativistic Ag bulk band structure (and using the irreducible
double-group representation), a significant spin-polarization (i.e., optical
pumping) should be expected at the 3--5~eV binding energy due to the
relativistic selection rules on bands with $\Lambda_{4,5}$ vs $\Lambda_6$
symmetry \cite{KuchRPP2001} (Figure~\ref{circular_dichroism}, bottom). We suggest that our low-symmetry situation might
render the changes in spin polarization detectable as intensity changes. We look
forward to detailed calculations to determine the particular physical origin of
our observed CDID signal. For now, we note that an analytical study \cite{Henk1996}
already indicated the possibility of observing circular dichroism due to
spin-orbit coupling under non-normal light incidence and normal emission on a
three-fold symmetric surface. (Our experimental spectra were acquired from a
single Ru terrace and a single Ag thickness, so the local substrate symmetry is
three-fold instead of six-fold \cite{de_la_figuera_determiningstructure_2006}.) Our choice of the azimuthal
direction of the light beam to be 11$^\circ$ off a compact direction on the Ru(0001) surface was ad hoc. As for the early CDAD work \cite{oepen_spin-dependent_1986}, we do not know how the strength of the CD signal varies as a function of azimuth. A systematic measurement of this dependence would be interesting, as was done in later CDAD experiments \cite{garbe_spin-dependent_1989}.

In summary, we have measured photoemission spectra from Ag layers 1--3 ML
thick by means of the dispersive mode of a spectromicroscope that
allows simultaneous characterization of the films with low-energy electrons both
in diffraction and in imaging mode. We observed circular dichroism in valence
band photoemission in a configuration that breaks the symmetry by having the in-plane light direction not aligned with a symmetry
plane of the surface. We term this circular dichroism from a non-magnetic surface
at normal-emission circular dichroism in the incidence distribution of photons,
CDID.

\ack
This research was supported by the U.S. Department of Energy under
contracts No. DE-AC04-94AL85000 and DE-AC02-05CH11231 and by the Spanish
Ministry of Science and Innovation under Projects No.~MAT2009-14578-C03-01 and No.~FIS2007-64982.

\section{References}
\bibliographystyle{unsrt}
\bibliography{vbd_AgRu}

\end{document}